\documentstyle[12pt,axodraw]{article}

\catcode`@=12
\begin{document}
\thispagestyle{empty}
\begin{flushright} 
UCRHEP-T326\\ 
TIFR/TH/01-48\\
November 2001\
\end{flushright}
\vspace{0.5in}
\begin{center}
{\LARGE	\bf Anomalous Neutrino Interaction, Muon g-2,\\ 
and Atomic Parity Nonconservation\\}
\vspace{1.5in}
{\bf Ernest Ma$^a$ and D. P. Roy$^b$\\}
\vspace{0.2in}
{$^a$ \sl Physics Department, University of California, Riverside, 
California 92521, USA\\}
\vspace{0.1in}
{$^b$ \sl Tata Institute of Fundamental Research, Mumbai (Bombay) 400005, 
India\\}
\vspace{1.5in}
\end{center}
\begin{abstract}\
We propose a simple unified description of two recent precision 
measurements which suggest new physics beyond the Standard Model 
of particle interactions, i.e. the deviation of $\sin^2 \theta_W$ in deep 
inelastic neutrino-nucleon scattering and that of the anomalous magnetic 
moment of the muon.  Our proposal is also consistent with a third precision 
measurement, i.e. that of parity nonconservation in atomic Cesium, which 
agrees with the Standard Model.
\end{abstract}
\newpage
\baselineskip 24pt

The minimal Standard Model (SM) of particle interactions is consistent with 
all present experimental data with only a few possible exceptions.  
One such is a recent measurement \cite{nutev} of the electroweak 
parameter $\sin^2 \theta_W$ from $\nu_\mu$ and $\bar \nu_\mu$ interactions 
with nucleons, which claims a three-standard-deviation departure from the SM 
prediction.  Another is the measurement \cite{g-2} of the 
anomalous magnetic moment of the muon, which originally claimed a value 
higher than the SM prediction by 2.6 standard deviations \cite{qcd}, but 
is now revised down to only 1.6$\sigma$ after a theoretical sign error has 
been corrected \cite{cor}.  A third 
important constraint comes from the measurement \cite{apv} of parity 
nonconservation in atomic Cesium, which was thought to be in disagreement 
with the SM, but subsequent improved theoretical calculations \cite{improved} 
have shown it to be in good agreement.  In addition, the phenomonena 
of neutrino oscillations are now well-established \cite{atm,solar} which 
suggest strongly that neutrinos have mass and mix with one another.

In this paper we propose a simple unified description of all the above 
effects by extending the SM to include the gauge symmetry $L_\mu - 
L_\tau$ \cite{hjlv}.  The relevance of this symmetry to the muon $g-2$ value 
and neutrino mass has been discussed by us in a previous paper \cite{maroy,
others}.  Here we focus on how it can also explain the NuTeV result 
\cite{nutev} and its other possible experimental consequences.

Our model assumes the anomaly-free gauge symmetry $U(1)_X$ with gauge boson 
$X$ which couples to $(\nu_\mu,\mu)_L$, $\mu_R$ with charge $+1$ and to 
$(\nu_\tau,\tau)_L$, $\tau_R$ with charge $-1$, but not to any other fermion. 
This means that it has the contribution
\begin{equation}
\Delta a_\mu = {g_X^2 m_\mu^2 \over 12 \pi^2 M_X^2}
\end{equation}
to the muon anomalous magnetic moment.  It also contributes to $\nu_\mu$ and 
$\bar \nu_\mu$ interactions, but since $X$ does not couple to quarks, the 
NuTeV result \cite{nutev} is only affected if $X$ mixes with the $Z$ boson 
of the SM.  This also applies to atomic parity nonconservation.

In our previous paper \cite{maroy}, we assume for simplicity that $X-Z$ 
mixing is zero by 
the imposition of an interchange symmetry in the Higgs sector, but we also 
mention that this symmetry cannot be maintained for the entire theory, so that 
a small deviation is to be expected.  This small deviation (corresponding to 
a mixing angle of order $10^{-3}$) turns out to be just what is needed to 
explain the NuTeV result, as shown below.

The Higgs sector of our model consists of three doublets: $\Phi = (\phi^+, 
\phi^0)$ with charge 0 and $\eta_{1,2} = (\eta^+_{1,2},\eta^0_{1,2})$ with 
charge $\pm 1$ under $U(1)_X$.  The mass matrix spanning $X$ and $Z$ is then 
given by
\begin{equation}
{\cal M}^2_{XZ} = \left[ \begin{array} {c@{\quad}c} 2 g_X^2 (v_1^2 + v_2^2) 
& g_X g_Z (v_1^2 - v_2^2) \\ g_X g_Z (v_1^2 - v_2^2) & (g_Z^2/2) (v_0^2 + 
v_1^2 + v_2^2) \end{array} \right],
\end{equation}
where $v_0 \equiv \langle \phi^0 \rangle$ and $v_{1,2}^2 \equiv \langle 
\eta^0_{1,2} \rangle$ with $v_0^2 + v_1^2 + v_2^2 = (2 \sqrt 2 G_F)^{-1}$. 
Assuming that $v_1 \simeq v_2$ so that the $X-Z$ mixing is small, we then 
have
\begin{equation}
M_Z^2 \simeq {1 \over 2} g_Z^2 (v_0^2 + 2 v_1^2), ~~~ M_X^2 \simeq 
4 g_X^2 v_1^2,
\end{equation}
with the $X-Z$ mixing angle given by
\begin{equation}
\sin \theta \simeq {g_X g_X (v_1^2 - v_2^2) \over M_X^2 - M_Z^2}.
\end{equation}

The effective $\nu_\mu$ and $\bar \nu_\mu$ interactions with quarks has 
the same structure as the SM, but the effective strength is changed from 
$g_Z^2/M_Z^2$ to
\begin{eqnarray}
&& g_Z^2 \left( {\cos^2 \theta \over M_Z^2} + {\sin^2 \theta \over M_X^2} 
\right) - 2 g_X g_Z \sin \theta \cos \theta \left( {1 \over M_Z^2} - {1 \over 
M_X^2} \right) \nonumber \\ && \simeq {g_Z^2 \over M_Z^2} \left[ 1 + 
{2 g_X \over g_Z} \left( {M_Z^2 \over M_X^2} - 1 \right) \sin \theta \right] 
\equiv {g_Z^2 \over M_Z^2} \rho_\mu.
\end{eqnarray}
Note that the factor of 2 in the $\sin \theta$ term comes from the fact that 
$X$ couples to $\nu_\mu$ with strength 1 whereas $Z$ couples to $\nu_\mu$ 
with strength 1/2 $(=I_3)$.

In the NuTeV analysis, if $\rho_\mu = 1$ is assumed, then $\sin^2 \theta_W 
= 0.2277 \pm 0.0013 \pm 0.0009$, which deviates from the SM prediction of 
$0.2227 \pm 0.00037$ by approximately $3\sigma$.  On the other hand, if a 
simultaneous fit to both $\rho_\mu$ and $\sin^2 \theta_W$ is made, they 
obtain
\begin{equation}
\rho_\mu = 0.9983 \pm 0.0040, ~~~ \sin^2 \theta_W = 0.2265 \pm 0.0031,
\end{equation}
with a correlation coefficient of 0.85 between the two parameters.  They 
then suggest that one but not both of them may be consistent with SM 
expectations.  Here we choose to consider the deviation of the NuTeV result 
as being due to $\rho_\mu$.

The NuTeV analysis also makes a two-parameter fit in terms of the isoscalar 
combinations of the effective neutral-current quark couplings, resulting in
\begin{equation}
(g_L^{eff})^2 = 0.3005 \pm 0.0014, ~~~ (g_R^{eff})^2 = 0.0310 \pm 0.0011,
\end{equation}
with a negligibly small correlation coefficient, whereas the SM 
predictions are
\begin{equation}
(g_L^{eff})^2_{SM} = 0.3042, ~~~ (g_R^{eff})^2_{SM} = 0.0301.
\end{equation}
Now if we take for example $\rho_\mu = 0.9962$, then the above two values 
become $(g_L^{eff})^2 = 0.3019$ and $(g_R^{eff})^2 = 0.0299$, placing them 
both within $1\sigma$ of the experimental measurements.

In atomic parity nonconservation, because $X$ does not couple to electrons, 
we have
\begin{equation}
\rho_e = \cos^2 \theta + \sin^2 \theta \left( {M_Z^2 \over M_X^2} \right) 
\simeq 1
\end{equation}
to a very good approximation.  Thus there should be no deviation from the 
SM, in agreement with experiment.

From Eq.~(5) we obtain
\begin{equation}
\sin \theta = (\rho_\mu - 1) \left( {g_Z \over 2 g_X} \right) \left( {M_X^2 
\over M_Z^2 - M_X^2} \right),
\end{equation}
which is of order $10^{-3}$ for $\rho_\mu = 0.9962$. 
This will affect precision data at the $Z$ resonance in the following way. 
First, the observed resonance is of course the physical $Z$ boson which 
has a small $X$ component.  However, since $X$ does not couple to electrons, 
the production of $Z$ is only suppressed by $\cos^2 \theta$ which is 
indistinguishable from 1.  The decay of $Z$ to most fermions is also 
unaffected because the suppression factor is again just $\cos^2 \theta$. 
The exceptions are $Z \to \mu^+ \mu^-, ~ \bar \nu_\mu \nu_\mu, ~ \tau^+ 
\tau^-, ~ \bar \nu_\tau \nu_\tau$.  Their effective couplings are
\begin{eqnarray}
\mu &:& g_V = -{1 \over 2} + 2 \sin^2 \theta_W - 2 \left( {g_X \over g_Z} 
\right) \sin \theta, ~~~ g_A = -{1 \over 2}, \\ 
\nu_\mu &:& g_V = {1 \over 2} - 2 \left( {g_X \over g_Z} \right) \sin \theta, 
~~~ g_A = {1 \over 2} - 2 \left( {g_X \over g_Z} \right) \sin \theta, \\ 
\tau &:& g_V = -{1 \over 2} + 2 \sin^2 \theta_W + 2 \left( {g_X \over g_Z} 
\right) \sin \theta, ~~~ g_A = -{1 \over 2}, \\ 
\nu_\tau &:& g_V = {1 \over 2} + 2 \left( {g_X \over g_Z} \right) \sin \theta, 
~~~ g_A = {1 \over 2} + 2 \left( {g_X \over g_Z} \right) \sin \theta.
\end{eqnarray}

Precision measurements of $Z$ couplings at LEP-I give \cite{euro}
\begin{equation}
g^\mu_V = -0.0359 \pm 0.0033, ~~~ g^\tau_V = -0.0366 \pm 0.0014,
\end{equation}
where the smaller error on $g^\tau_V$ is due to the use of $\tau$
polarization along with the forward-backward asymmetry.  Thus 
\begin{equation}
g^\tau_V - g^\mu_V = 4 (g_X/g_Z) \sin \theta  = -0.0007 \pm 0.0036, 
\end{equation}
adding the two errors in quadrature.  Consider now Eq.~(10) with the 
more conservative choice
\begin{equation}
\rho_\mu = 0.9976
\end{equation}
which is within $1.6\sigma$ of the NuTeV measurement of $(g_L^{eff})^2$. 
Comparing it to Eq.~(16), we then obtain the following $2\sigma$ bounds 
on $M_X$:
\begin{equation}
M_X < 72~{\rm GeV} ~{\rm or}~ M_X > 178~{\rm GeV}.
\end{equation}

A lower bound on $M_X$ as a function of $g_X$ is also available from LEP-I 
data on $Z$ decay into the 4-muon final state via $Z \rightarrow \mu^+\mu^- 
X$ \cite{maroy}.  For example, if $g_X = 0.2$, then $M_X > 58$ GeV.  
Furthermore, Eq.~(3) requires
\begin{equation}
g_X > {g_Z M_X \over 2 M_Z}.
\end{equation}
In Figure 1 we show the above lower limit on $g_X$ as well as the $2\sigma$ 
upper limits on $g_X$ as functions of $M_X$ from $Z \to \mu^+ \mu^- X$ 
decay and the difference of the $Z \to e^+ e^-$ and $Z \to \mu^+ \mu^-$ 
partial widths as the result of the $X$ radiative contribution.  Details 
are provided in Ref.~[9].  The $Z$ decay limit essentially rules out 
$M_X < 60$ GeV.  The analogous process $e^+ e^- \to \mu^+ \mu^- X$ at 
LEP-II does not improve this bound, as already shown \cite{maroy}.  
Thus we conclude that $M_X$ between 60 and 72 GeV is still allowed, 
but perhaps $M_X > 178$ GeV is more likely.

Going back to Eq.~(1) for the muon $g-2$ discrepancy, we note that 
there is a theoretical \underline {lower} bound \cite{maroy} of $1.56 \times 
10^{-9}$ in this model, whereas the corrected \cite{cor} range of the 
experimental discrepancy is $2.65 \pm 1.65 \times 10^{-9}$.  This is 
entirely consistent with the low $M_X$ solution, while in the case of the 
high $M_X$ solution, the maximum deviation we get is $2.7 \times 10^{-9}$.  In 
either case, the $X$ boson signal will be too small to be observable at the 
Fermilab Tevatron, but will be clearly visible at the CERN LHC \cite{maroy} 
via the associated production processes $u \bar u (d \bar d) \to \mu \mu X$ 
and $u \bar d (d \bar u) \to \mu \nu X$.  At a future muon collider, $X$ 
would be copiously produced, especially if it turns out to be light.

To obtain naturally small Majorana neutrino masses, we may add one heavy 
neutral fermion singlet $N_R$ with $U(1)_X$ charge 0 as in our previous 
paper, but then an extra charged scalar boson $\zeta^+$ with charge +1 
is needed there to get a second neutrino mass term, i.e. $\nu_e \nu_\tau$, 
radiatively.  A possible alternative is to add two $N_R$'s.  One is 
assumed to couple only to a linear combination of $(\nu_\mu \eta_2^0 -\mu_L 
\eta^+_2)$ and $(\nu_\tau \eta^0_1 - \tau_L \eta_1^+)$, and the other to 
$(\nu_e \phi^0 - e_L \phi^+)$ as well.  Using the canonical seesaw 
mechanism \cite{seesaw}, this structure allows for the appearance of two 
massive neutrinos: one is predominantly a mixture of $\nu_\mu$ and $\nu_\tau$, 
the other is a linear combination of $\nu_e$ and the orthogonal $\nu_\mu - 
\nu_\tau$ mixture.  This may then lead to a consistent pattern of neutrino 
masses and mixing for explaining the present atmospheric \cite{atm} and solar 
\cite{solar} neutrino data.

The interchange symmetry $\eta_1 \leftrightarrow \eta_2$ in the Higgs 
sector allows us to assume $v_1 = v_2$, but this cannot be maintained 
for the entire theory.  If we try to extend this to the gauge sector, then 
$\mu \leftrightarrow \tau$ is implied.  Hence $m_\mu \neq m_\tau$ in the 
Yukawa sector would break this symmetry.  However, the size of this breaking 
is only of order 
$(m_\tau^2 - m_\mu^2)/v_0^2$ which is smaller than what we require for 
$\sin \theta$.  In other words, $X-Z$ mixing of order $10^{-3}$ is a very 
reasonable value.

In conclusion we have shown in this paper how the gauge symmetry $L_\mu - 
L_\tau$ (as realized specifically by us in a previous paper \cite{maroy}) 
explains 
naturally the recent NuTeV result \cite{nutev} on the possible deviation 
from the Standard Model in $\nu_\mu$ and $\bar \nu_\mu$ scattering with 
nucleons.  Our proposal also explains the possible discrepancy in the 
recent measurement \cite{g-2} of the anomalous magnetic moment of the 
muon.  It further explains why there is no deviation from the Standard Model 
in atomic parity nonconservation \cite{apv}.  Our model is constrained by 
the precision measurements of $Z \to \mu^+ \mu^-$ and $Z \to \tau^+ \tau^-$, 
from which we predict that the new gauge boson $X$ is likely to have a mass 
between 60 and 72 GeV, or be heavier than 178 GeV.  As such, our model is 
verifiable experimentally in the future at the LHC.

We thank S.~N.~Ganguli and A.~Gurtu for discussions on the LEP data.
This work was supported in part by the U.~S.~Department of Energy under 
Grant No.~DE-FG03-94ER40837.

\bibliographystyle{unsrt}

\newpage

\begin{figure}
\begin{center}
\vspace*{3.5in}
\includegraphics{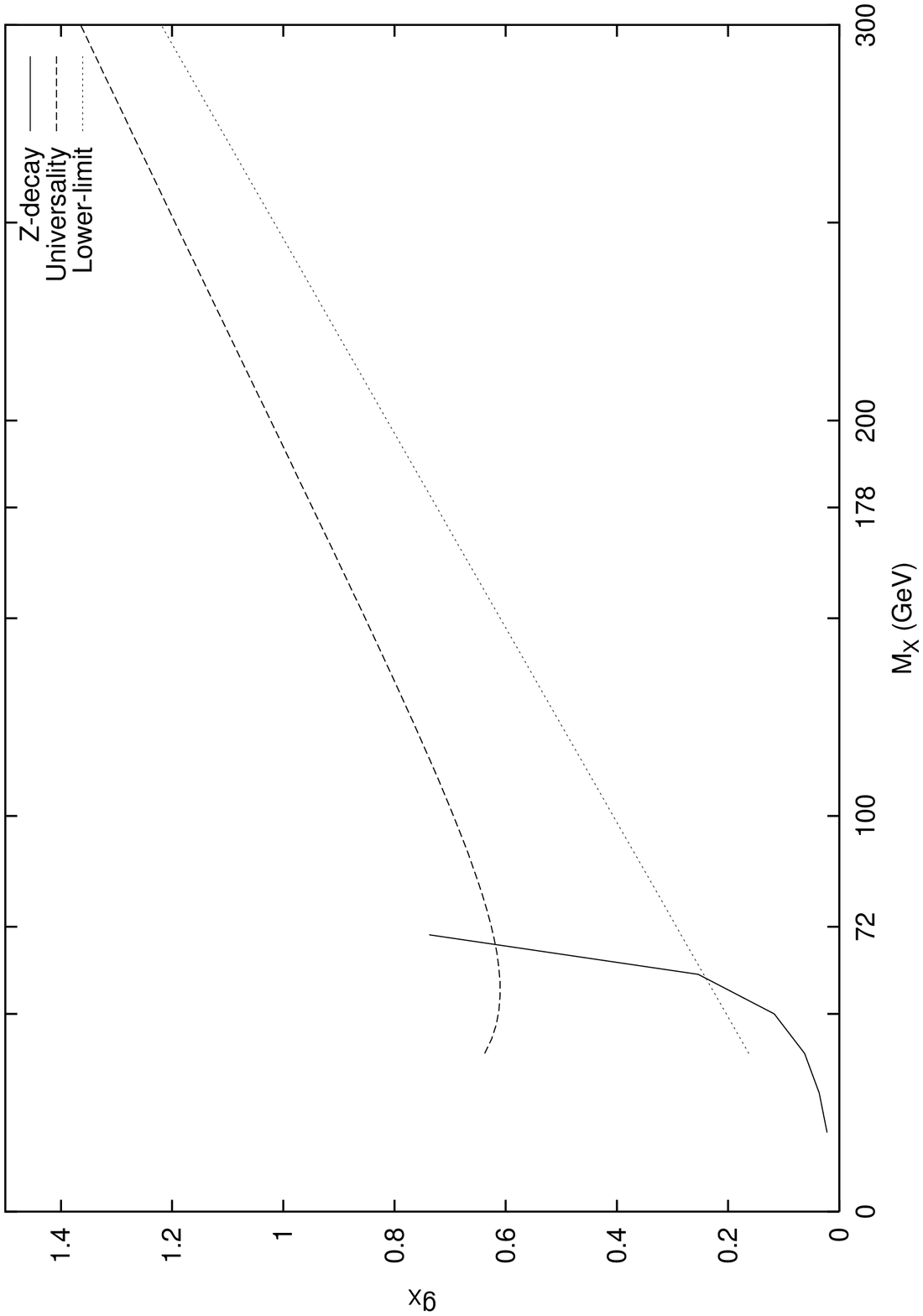}
\end{center}
\caption{The predicted lower limit of the $X$ boson coupling shown
along with the LEP-I upper limits from $Z \rightarrow \mu^+\mu^- X$
decay and the universality relation between the $Z \rightarrow e^+e^-$
and $\mu^+\mu^-$ partial widths.  The $X$ mass ranges of interest to
the NuTeV anomaly are $M_X = 60 - 72$ GeV or $M_X > 178$ GeV.}
\end{figure}

\end{document}